\documentclass[preprint,showpacs,showkeys,preprintnumbers,amsmath,amssymb]{revtex4}
\begin{document}
\title{Rayleigh's collapsing time of a spherical cavity: the $\Delta$-factor}
\author{J. A. S. Lima}
\email{limajas@astro.iag.usp.br} \affiliation{Universidade de S\~ao
Paulo, Instituto de Astronomia, Geof\'\i sica e Ci\^encias
Atmosf\'ericas \\ Rua do Mat\~ao, 1226, CEP 05508-900, S\~ao Paulo,
SP, Brazil}
\author{F. E. M. da Silveira}
\email{feugenio@if.usp.br} \affiliation{Universidade de S\~ao Paulo,
Instituto de F\'\i sica\\ Caixa Postal 66318, CEP 05315-970, S\~ao
Paulo, SP, Brazil}
\date{\today}
\begin{abstract}
New corrections to the equation of motion and total collapsing time
of an empty spherical cavity immersed in an infinite incompressible
medium are proposed on the assumption of a non-uniform density. The
dimensionless number quantifying the corrections with respect to the
standard Rayleigh results (coined the $\Delta$-factor) is fully
independent of other possible contributions like surface tension and
viscous terms. The $\Delta$-factor effect advocated here can be seen
as a direct consequence of a mass-shell non-trivial solution to the
continuity equation. The consistency of the corrections with respect
to the Bernoulli theorem and some physical consequences in the
framework of the Rayleigh-Plesset equation are also discussed.
\end{abstract}
 \maketitle
\section{introduction}
Ninety years ago, Lord Rayleigh published his seminal article on the
problem of determining the collapsing time of an empty spherical
cavity, starting from an arbitrary radius $a$, suddenly formed in
the bulk of an infinite medium \cite{ray1917}. Later on, that paper
became the predecessor of a very active area of research,
collectively referred to as bubble dy\-na\-mics, and, as such, it
can be regarded as a kind of paradigm to a vast class of spherically
symmetric one-dimensional non-steady flows. To quote some examples,
Rayleigh's work is closely related to a large number of applications
in the development of engineering devices \cite{ENG},
sonoluminescence \cite{SN}, and, more generally, with the important
problem of cavitation \cite{cavity}. Nowadays, that classical
approach is the starting point for a large number of investigations,
based on the medium properties therein neglected, like surface
tension, fluid viscosity, heat transfer, acoustic cavitation, bubble
relaxation, diffusive terms, shock-wave induced collapse, chaos, and
many others [5-9](see also \cite{revPP,revBrenner1} for reviews on
such subjects).

Rayleigh achieved his celebrated result on the basis of a few number
of hypothesis, thereby capturing the essence of the physics
contained in the collapsing cavity. First, he considered the fluid
to be incompressible, that is, to have a divergenceless flow. By
adopting a spherical coordinate system,  concentric with the cavity,
this means that the negative radial component of the velocity field
$v\left(r,t\right)$ may be expressed as
\begin{eqnarray}
v=\frac{R^2\dot R}{r^2}, \label{1}
\end{eqnarray}
where $R\left(t\right)$ denotes the instantaneous radius of the
cavity, and the dot represents total time derivative. Further, by
assuming a uniform fluid density $\rho=\rho_{\infty}$ = constant,
and applying the work-kinetic energy theorem, he was able to relate
the collapsing rapidity of the cavity's boundary to its
instantaneous radius through
\begin{eqnarray}
\dot{R}^2=\frac{2p_\infty}{3\rho_{\infty}}\left(\frac{a^3}{R^3}-1\right),
\label{2}
\end{eqnarray}
where $a$ is the initial radius of the cavity and $p_\infty$ stands
for the constant fluid pressure at infinity. In the above
expression, the instantaneous pressure on the cavity's boundary is
presumed to vanish, once effects provoked by surface tension and
viscous terms are neglected. Finally, in order to determine the
total collapsing time $\tau_{\rm R}$ of the cavity, Rayleigh
integrated Eq. (\ref{2}) with respect to $R$ from the initial radius
to zero, thereby obtaining
\begin{eqnarray}
\tau_{\rm R}=a\sqrt{\frac{3\pi\rho_\infty}{2p_\infty}}
\frac{\Gamma\left(5/6\right)}{\Gamma\left(1/3\right)}, \label{3}
\end{eqnarray}
where $\Gamma\left(m/n\right)$ is the gamma function. Here, we
notice that Eq. (\ref{2}) is easily shown to be a particular first
integral of
\begin{eqnarray}
R\ddot{R}+\frac{3}{2}\dot{R}^2+\frac{p_\infty}{\rho_\infty}=0,
\label{4}
\end{eqnarray}
which, from now on, will be referred to as Rayleigh's equation of
motion.

In what follows, we do not try to include any additional
phenomenological effect into Rayleigh's bubble dynamics, as the ones
mentioned in the above quoted references. Instead, we seek
fundamental modifications of his results by relaxing the hypothesis
of uniform density assumed by Rayleigh, as well as in some of their
important developments like in the so-called Rayleigh-Plesset
equation \cite{PlessetE}.

The article is organized as follows. In section II, we derive an
analytical, non-trivial solution to the continuity equation, which
is consistent with the hypothesis of a divergenceless flow. In
section III, the $\Delta$-factor is defined, leading to an extension
of Rayleigh's equation of motion and a correction to the total
collapsing time of the cavity. In sections IV and V, we discuss the
consistency of the corrections with the Bernoulli theorem and some
physical consequences in the framework of the so-called
Rayleigh-Plesset equation, respectively. Finally, in the conclusion
section, the main results are summarized.

\section{mass-shell solution}

For a spherically symmetric one-dimensional non-steady flow, the
fluid density $\rho\left(r,t\right)$ is demanded to satisfy the
continuity equation in the form
\begin{eqnarray}
\frac{\partial\rho}{\partial t}
+\frac{1}{r^2}\frac{\partial}{\partial r}\left(r^2\rho v\right)=0.
\label{5}
\end{eqnarray}
Now, once Eq. (\ref{1}) is assumed to hold true, and changing
coordinates from time $t$ to the instantaneous radius
$R\left(t\right)$, Eq. (\ref{5}) readily leads to
\begin{eqnarray}
\frac{1}{R^2}\frac{\partial\rho}{\partial R}
+\frac{1}{r^2}\frac{\partial\rho}{\partial r}=0. \label{6}
\end{eqnarray}
As it appears, one is tempted to take $\rho$ = constant as the
solution to the above equation which is consistent with the
divergenceless flow. However, by considering the region $a\leq r\leq
b$ of the fluid, a more general solution to that equation may be
written as
\begin{eqnarray}
\rho\left(r,R\right)=\frac{\rho_b-\rho_a{\rm e}^{-\lambda}}{1-{\rm
e}^{-\lambda}} +\frac{\rho_a-\rho_b}{1-{\rm e}^{-\lambda}}
\exp\left[-\left(\frac{r^3}{a^3}-\frac{R^3}{a^3}\right)\right],
\label{7}
\end{eqnarray}
where we have chosen $\rho_a=\rho\left(a,a\right)$ and
$\rho_b=\rho\left(b,a\right)$ as boundary conditions at the initial
instant, whilst the dimensionless boundary factor $\lambda$ has been
defined as
\begin{eqnarray}
\lambda=\frac{b^3}{a^3}-1. \label{8}
\end{eqnarray}
Note that Eq. (\ref{7}) actually describes a vast collection of
radially stratified mass-shells at some fixed instant.
Notwithstanding, since we are describing a finite class of
solutions, Rayleigh's one may be contained into this as a particular
case. Indeed, by identifying $a$ to the initial radius of the
cavity, and taking the limit $b\rightarrow\infty$, from Eq.
(\ref{7}), it directly follows that
\begin{eqnarray}
\rho\left(r,R\right)=\rho_\infty+\left(\rho_0-\rho_\infty\right)
\exp\left[-\left(\frac{r^3}{a^3}-\frac{R^3}{a^3}\right)\right],
\label{9}
\end{eqnarray}
where we have renamed $\rho_0=\rho_a$ and
$\rho_\infty=\rho_{b\rightarrow\infty}$. In that case, clearly,
$\rho_0$ denotes the instantaneous density on the cavity boundary
and $\rho_\infty$ represents the constant density at infinity. Of
course, by putting $\rho_0=\rho_\infty$ = constant, Eq. (\ref{9})
trivially recovers the solution adopted by Rayleigh. In principle,
when a spherical bubble is formed in an uniform medium, one may
expect a variation of the density with the radial coordinate. Since
the mass is conserved and a hole has been somewhat created, the mass
must be redistributed (compressed) close to the frontier of the
cavity. In this way, the density near the cavity boundary becomes
greater than far from the wall, and, solution (\ref{9}), says that
such a decay is actually very fast ($\propto e^{-\frac{r^3}{a^3}}$)
for a divergenceless flow.

Here, for further purposes, we rewrite Eq. (\ref{9}) in the form
\begin{eqnarray}
\rho\left(x\right)=\rho_\infty\left(1+A{\rm e}^{-x}\right),
\label{10}
\end{eqnarray}
where we have changed coordinates from radius $r$ to the
dimensionless quantity $x$ through the transformation
\begin{eqnarray}
x=\frac{r^3}{a^3}-\frac{R^3}{a^3}, \label{11}
\end{eqnarray}
for some fixed $R\left(t\right)$, as well as defined the
dimensionless $A$-factor by the relation
\begin{eqnarray}
A=\frac{\rho_0}{\rho_\infty}-1. \label{12}
\end{eqnarray}

Given the above analytical possibility of a non-uniform fluid
density, as the only new modification with respect to Rayleigh's
original formulation, and since the assumption of a vanishing
instantaneous pressure on the cavity's boundary still holds, one can
anticipate that any correction to the total collapsing time of the
cavity must involve only some dimensionless combination of both
$\rho_0$ and $\rho_\infty$. Let us now carefully examine the
proposed generalization of the original Rayleigh equation of motion,
which shall show consistency with our non-uniform density solution,
as well as lead to a correction to the standard total collapsing
time of the cavity.
\section{generalized rayleigh's equation: the $\Delta$-factor}
The fluid pressure $p\left(r,t\right)$ is required to satisfy
Euler's equation in the form
\begin{eqnarray}
\frac{\partial v}{\partial t}+v\frac{\partial v}{\partial r}
+\frac{1}{\rho}\frac{\partial p}{\partial r}=0, \label{13}
\end{eqnarray}
for a spherically symmetric one-dimensional non-steady flow. Once
Eqs. (\ref{1}) and (\ref{10}) are assumed to hold true and taking
into account Eq. (\ref{11}), it proves convenient to introduce the
dimensionless functional
\begin{eqnarray}
y\left(x\right)=\frac{p\left(x\right)}{p_\infty}, \label{14}
\end{eqnarray}
for some fixed instant $t$, or, equivalently, for some fixed
instantaneous radius $R\left(t\right)$. Now, by integrating Eq.
(\ref{13}) from the instantaneous boundary of the cavity to
infinity, provided the instantaneous pressure on the cavity's
boundary vanishes, we obtain the extended Rayleigh equation of
motion
\begin{eqnarray}
R\ddot{R}+\frac{3}{2}\dot{R}^2+\Delta\frac{p_\infty}{\rho_\infty}=0,
\label{15}
\end{eqnarray}
where, from Eqs. (\ref{10}) and (\ref{14}), the time independent
dimensionless $\Delta$-factor is defined by
\begin{eqnarray}
\Delta=\int_0^\infty\frac{1}{1+A{\rm e}^{-x}}\frac{\partial
y}{\partial x}dx. \label{16}
\end{eqnarray}
As one may check, by applying the initial condition on the surface
of the empty cavity [$\dot{R}\left(0\right)=0$ for $R=a$], a first
integral of Eq. (\ref{15}) reads
\begin{eqnarray}
\dot{R}^2=\frac{2\Delta
p_\infty}{3\rho_{\infty}}\left(\frac{a^3}{R^3}-1\right). \label{17}
\end{eqnarray}
As a result, the corrected total collapsing time $\tau_\Delta$ of
the cavity can be written as
\begin{eqnarray}
\tau_\Delta=\frac{\tau_{\rm R}}{\sqrt{\Delta}}, \label{18}
\end{eqnarray}
which could be anticipated from the equation of motion (e.g., by
replacing  $p_\infty\rightarrow\Delta p_{\infty}$).

All the above results should be compared  with  Eqs. (\ref{2}),
(\ref{3}) and  (\ref{4}) of the introduction. In the limit
$A\rightarrow 0$, that is, by putting $\rho_0=\rho_\infty$ =
constant, we see that Eq. (\ref{16}) implies $\Delta=1$, thereby
recovering the respective Rayleigh results. Naturally, in order to
quantify an arbitrary $\Delta$-contribution, the integral given by
Eq. (\ref{16}) must somewhat be evaluated. Notwithstanding, we
discuss now a simple estimate of
$\Delta\left(\rho_0/\rho_{\infty}\right)$, by assuming that $A$ is
sufficiently large, that is, $\rho_0\gg\rho_\infty$. An independent
derivation based on the Bernoulli equation can be seen at the end of
the next section.

To begin with, we observe that the functionals $\rho\left(x\right)$
and $p\left(x\right)$ over the domain of integration ($0<x<\infty$)
suggest the definition of a large dimensionless quantity $x_\ast$,
defined as the value $x$ takes such that
\begin{eqnarray}
y\left(x_\ast\right)=\frac{1}{2}. \label{19}
\end{eqnarray}
Next, we separate the integral in Eq. (\ref{16}) in two parts,
namely, the first over the interval $0<x<x_\ast$, and the second one
for $x_\ast<x<\infty$. Since $A$ is assumed to be large enough and
the cavity very small as compared to the infinite medium, we may
neglect the functional dependence of the integrating factor on $x$
throughout the first interval, whilst keep it through the second
one. However, since $x$ is also large throughout the second
interval, the whole integrating factor may be fairly well
approximated to unity over there. As a consequence, we may
approximate our Eq. (\ref{16}) by
\begin{eqnarray}
\Delta=\frac{1}{1+A}\int_{0}^{x_\ast}\frac{\partial y}{\partial x}dx
+\int_{x_\ast}^{\infty}\frac{\partial y}{\partial x}dx, \label{20}
\end{eqnarray}
from which it immediately follows that
\begin{eqnarray}
\Delta=\frac{\rho_0+\rho_\infty}{2\rho_0}. \label{21}
\end{eqnarray}
Note that, as long as the instantaneous density on the cavity's
boundary severely exceeds the constant density at infinity, $\Delta$
approaches its lowest possible value, namely
\begin{eqnarray}
\Delta_{\rm min}=\frac{1}{2}. \label{22}
\end{eqnarray}
As a consequence, the corrected total collapsing time of the cavity
approaches its largest possible value, namely
\begin{eqnarray}
\tau_{\rm max}=\tau_{\rm R}\sqrt{2}. \label{23}
\end{eqnarray}

Let us now investigate the consistency of the $\Delta$-factor as
given by Eqs. (\ref{16}) and (\ref{21}) with Bernoulli's theorem. As
we shall see, the foregoing independent argument leads to the same
$\Delta$-factor and Rayleigh's type equation of motion.
\section{$\Delta$-factor and the Bernoulli theorem}
The fluid motion represented by Eq. (\ref{1}) implies
$\nabla\times\vec{v}=0$, i. e., the velocity field in the Rayleigh
dynamics yields a potential flow. As a consequence, we can
immediately write $\vec{v}=\nabla\phi$, where the hydrodynamical
potential $\phi\left(r,t\right)$ may be chosen to be
\begin{eqnarray}
\phi=-\frac{R^2\dot{R}}{r}. \label{24}
\end{eqnarray}
Now, for an isentropic flow, viz., on the assumption of a uniform
specific (per unit mass) entropy $s$ throughout the fluid, a
generalized form of Bernoulli's theorem shall be written as
\cite{Landau, Batchelor}
\begin{eqnarray}
\frac{\partial\phi}{\partial
t}+\frac{1}{2}v^2+\omega=f\left(t\right), \label{25}
\end{eqnarray}
where $\omega\left(r,t\right)$ denotes the fluid enthalpy, and
$f\left(t\right)$ represents some arbitrary function of time. At
this point, we can, simultaneously, evaluate the left-hand side of
Eq. (\ref{25}) on the cavity's boundary, that is, at $r=R$, and,
consistently with Eqs. (\ref{1}) and (\ref{24}), which ensure that
both $v\left(r,t\right)$ and $\phi\left(r,t\right)$ vanish in the
limit $r\rightarrow\infty$, choose $f\left(t\right)$ to be the
constant value $\omega\left(r,t\right)$ takes at infinity. In that
case, once $\omega\left(r,t\right)$ is related to the specific
internal energy $\epsilon\left(r,t\right)$ of the fluid through
\begin{eqnarray}
\omega=\epsilon+\frac{p}{\rho}, \label{26}
\end{eqnarray}
by assuming a vanishing instantaneous pressure on the cavity's
boundary, Eqs. (\ref{25}) and (\ref{26}) lead to
\begin{eqnarray}
R\ddot{R}+\frac{3}{2}\dot{R}^2+\frac{p_\infty}{\rho_\infty}
+\left(\epsilon_\infty-\epsilon_0\right)=0, \label{27}
\end{eqnarray}
where $\epsilon_0$ and $\epsilon_\infty$ denote the constant values
$\epsilon\left(r,t\right)$ takes at the cavity's boundary and at
infinity, respectively. It thus follows that any correction to the
original Rayleigh's equation of motion may only come from the
parenthesis in Eq. (\ref{27}). In order to evaluate it, we shall
make use of the well-known thermodynamical relation
\begin{eqnarray}
d\epsilon=Tds+p\frac{d\rho}{\rho^2}, \label{28}
\end{eqnarray}
where $T\left(r,t\right)$ denotes the thermodynamical temperature
throughout the fluid. However, since we have an isentropic flow
($ds=0$), we can rewrite Eq. (\ref{27}) as
\begin{eqnarray}
R\ddot{R}+\frac{3}{2}\dot{R}^2+\frac{p_\infty}{\rho_\infty}
+\int_{\rho_0}^{\rho_\infty}p\frac{d\rho}{\rho^2}=0. \label{29}
\end{eqnarray}
It is a trivial matter to show that the above equation is equivalent
to the previously derived form of the generalized Rayleigh's
equation of motion, Eq. (\ref{15}), provided the integral in Eq.
(\ref{29}) is carried out by parts,
\begin{eqnarray}
\int_{\rho_0}^{\rho_\infty}p\frac{d\rho}{\rho^2}=-\frac{p_\infty}{\rho_\infty}
+\int_{R}^{\infty}\frac{1}{\rho}\frac{\partial p}{\partial r}dr.
\label{30}
\end{eqnarray}
As a consequence, from Eq. (\ref{27}), we arrive at an independent
definition of the $\Delta$-factor with respect to the one given in
Eq. (\ref{16}),
\begin{eqnarray}
\Delta=1+\frac{\rho_\infty}{p_\infty}\left(\epsilon_\infty-\epsilon_0\right)
=1+\frac{\rho_\infty}{p_\infty}\int_{\rho_0}^{\rho_\infty}p\frac{d\rho}{\rho^2}.
\label{31}
\end{eqnarray}

At this point, through a quite simple argument, consistently with a
non-uniform density solution, we show that the same estimate of the
$\Delta$-factor as given by Eq. (\ref{21}) can be achieved in the
framework of Bernoulli's theorem. The unique necessary assumption is
the monotonical growth of the fluid pressure throughout the
increasing radial coordinate concentric with the cavity. We start by
introducing an instantaneously averaged fluid pressure, $\bar{p}$
say, defined throughout the bulk of the infinite medium as
\begin{eqnarray}
\bar{p}=\lim_{N\rightarrow\infty}\frac{1}{N} \left[\delta p+2\delta
p+3\delta p +\ldots+\left(N-1\right)\delta p+p_\infty\right],
\label{32}
\end{eqnarray}
where $N$ denotes a very large number of mass-shells, on the
assumption that $p\left(\rho\right)$ on each radially stratified
layer. In principle, one should expect $p$ to be dependent on the
fluid entropy, as well. However, we have assumed the flow to be
isentropic. In this proposed scheme, two neighboring surfaces, no
matter if they stand close together or far apart throughout the
radial direction, are labeled by two different values of
$p\left(\rho\right)$, whose difference is exactly equal to a very
small number, $\delta p$, say.  Note that $p_\infty=N\delta p$,
whilst the pressure on the cavity's boundary presumably vanishes. In
addition, consistently with the assumption that $p_\infty$ =
constant, $\delta p\rightarrow0$ in the limit $N\rightarrow\infty$.
Now, the problem of evaluating the sum in Eq. (\ref{32}) is a
trivial one. The answer is
\begin{eqnarray}
\delta p+2\delta p+3\delta p+\ldots+\left(N-1\right)\delta
p+p_\infty =\frac{N\left(\delta p+p_\infty\right)}{2}. \label{33}
\end{eqnarray}
In other words, the sought instantaneously averaged fluid pressure,
actually, does not depend on time, it is a constant, and we can
legitimately approximate the pressure throughout the volume of the
fluid by
\begin{eqnarray}
\bar{p}=\frac{p_\infty}{2}. \label{34}
\end{eqnarray}
As one may check, by inserting Eq. (\ref{34}) into Eq. (\ref{31}),
our previous expression for $\Delta$ as given by Eq. (\ref{21}) is
recovered.
\section{$\Delta$-factor and Rayleigh-Plesset equation}
It is also interesting to discuss how the $\Delta$-factor (or the
variation of the density) may modify the description of other fluid
properties, from the point of view of bubble dynamics theory. In
this regard, an important analytical development was achieved by
Plesset thus leading to what is now often referred to as
Rayleigh-Plesset equation. That extended theory is remarkably simple
and, in particular, explains several features related to
single-bubble sonoluminescence experiments. In such an approach, the
fluid density is also constant but one assumes a gas-filled cavity
so that several boundary effects are taken into account as well. The
modified equation of motion reads \cite{revPP, PlessetE}
\begin{eqnarray}
R\ddot{R}+\frac{3}{2}\dot{R}^2+\frac{{p_\infty}-p\left(R\right)}{\rho_\infty}=0,
\label{35}
\end{eqnarray}
where $p(R)$ is the pressure of the liquid at the bubble boundary.
It includes terms like surface-tension, viscous stress and the
pressure in the bubble, $p_B$, and can be written as
\begin{eqnarray}
p\left(R\right)={p_{\rm B}}-\frac{2\sigma}{R}-{4\mu}\frac{\dot
R}{R}, \label{36}
\end{eqnarray}
where $\sigma$ is the surface-tension constant and $\mu$ is the
viscosity coefficient. From the above expression, the
Rayleigh-Plesset equation assumes the following form
\begin{eqnarray}
R\ddot{R}+\frac{3}{2}\dot{R}^2+\frac{1}{\rho_\infty}\left(p_\infty
+\frac{2\sigma}{R}+{4\mu}\frac{\dot{R}}{R}-p_{\rm B}\right)=0.
\label{37}
\end{eqnarray}
An important observable difference between Eqs. (\ref{37}) and
(\ref{4}) is the existence of an e\-qui\-li\-brium (unstable)
radius, $R_{\rm E}$, which is obtained by imposing the conditions
$\dot{R}=0$ and $\ddot{R}=0$ to be simultaneously fulfilled. For
further reference, it is given by
\begin{eqnarray}
R_{\rm E}=\frac{2\sigma}{p_{\rm B}-p_\infty},
\label{38}
\end{eqnarray}
which does not depend explicitly on the fluid density. Additional
properties of the Rayleigh-Plesset equation have been widely
discussed in the literature \cite{revPP,revBrenner1}.

In the above context, the interesting question is: how the
non-uniform density solution of section II will affect the
Rayleigh-Plesset equation? Naturally, one may also expect a
modification of the standard equilibrium radius.

Following the same approach presented in Section III, it is readily
checked that the Rayleigh-Plesset relation, Eq. (\ref{37}), must be
replaced by
\begin{eqnarray}
R\ddot{R}+\frac{3}{2}\dot{R}^2+\Delta\frac{p_\infty}{\rho_\infty}
+\frac{2\sigma}{\rho_0 R}+\frac{4\mu}{\rho_0}\frac{\dot R}{R}
-\frac{p_{\rm B}}{\rho_0}=0.
\label{39}
\end{eqnarray}
As one should expect, beyond the modification of the
$\Delta$-factor, all boundary terms are now divided by the density
of the liquid in the evolving bubble frontier ($\rho_0$). In
addition, the equilibrium conditions lead to a corrected equilibrium
radius of the bubble
%\begin{eqnarray}
%R_{\rm E}\left(\Delta\right)=\frac{2\sigma}
%{p_{\rm B}-\Delta p_\infty\frac{\rho_0}{\rho_\infty}},
%\label{40}
%\end{eqnarray}
\begin{eqnarray}
R_{\rm E}\left(\Delta\right)=\frac{2\sigma}
{p_{\rm B}-\Delta \left(\rho_0/\rho_\infty\right)p_\infty},
\label{40}
\end{eqnarray}
which depends explicitly on the boundary value of the density. It
is also worth notice that in the limit $\Delta=1$, that is, for
$\rho_0=\rho_\infty$ = constant, Eq. (\ref{40}) trivially recovers
the standard equilibrium radius of the bubble as predicted by the
standard Rayleigh-Plesset equation.
\section{conclusion}
In this work, we have discussed a new kind of contribution to
Rayleigh's equation of motion and the total collapsing time of an
empty spherical cavity. As we have seen, a correction term, herein
called the $\Delta$-factor, naturally appears if the fluid density
is not uniform, as assumed by many authors including Rayleigh
himself. Indeed, the $\Delta$-factor is a direct consequence of a
divergenceless spherically symmetric non-steady flow, whose fluid
density is an evolving function in space and time, as given by the
general solution derived in section II [see Eq. (\ref{9})].

The inclusion of the $\Delta$-factor as a new parameter into
Rayleigh's dynamics ne\-cessarily leads to non-negligible
corrections of a number of meaningful physical quantities, as the
collapsing time. We have also shown how the extension usually named
Rayleigh-Plesset equation is affected by the non-uniform time
varying density. In particular, a new equilibrium radius has also
been derived. Unlike the standard approach, it depends explicitly
on the fluid density at the boundary and far from the bubble. All
the results summarized above are analytical and have been justified
from first principles. It should also be stressed that even more general
formulations, as the one describing approximately the influence of the
liquid compressibility \cite{KM80}, may be affected by the correction
proposed here. A more detailed investigation of the $\Delta$-factor on
bubble dynamics will be discussed in a forthcoming communication.

\begin{acknowledgments}
This work has been partially supported by CNPq, Brazilian research
agency. JASL is also grateful to FAPESP No. 04/13668-0.
\end{acknowledgments}
{}
\end{document}